\documentstyle[preprint,eqsecnum,aps]{revtex}
\begin{document}
\draft

\preprint{ZTF--93/9--R}

\title{Covariant model of a quarkonium with the funnel potential}

\author{R. Horvat, D. Kekez, and D. Palle}
\address{Rudjer Bo\v{s}kovi\'{c} Institute,
         P.O.B. 1016, 41001 Zagreb, Croatia}
\author{D. Klabu\v{c}ar}
\address{Physics Department of Zagreb University,
         P.O.B. 162, 41001 Zagreb, Croatia}

\maketitle

\begin{abstract}
   The bound--state problem for the pion as a quarkonium with the funnel
(Coulomb--plus--linear) interaction is solved in a framework
that combines the bilocal approach to mesons with the covariant
generalization of the instantaneous--potential model.
The potential interaction leads
to dynamical breaking of chiral symmetry.
However, the Coulomb potential leads to ultraviolet divergences that must
be subtracted.
A careful choice of the renormalization prescription is needed in order
to get the correct chiral limit.
The mass, the lepton
decay constant of the pion, as well as the pion decay width in two photons
are calculated.
\end{abstract}

\narrowtext

\section{Introduction}
\label{INTRODUCTION}

   The activity in the field of relativistic bound states has recently
increased. References \cite{bicudo92,gross92,gross92b,mitra92b,williams93}
are just
some of examples. These papers were motivated, among other things,
by the desire to formulate a covariant
treatment for quarks interacting through a potential which would hopefully
mimic nonperturbative QCD.

   Our work is a continuation of the line of research where interquark
interactions were modeled by the instantaneous potential.
For the case of light quarks, Refs.~\cite{leyaouanc84,adler84,trzupek89}
may serve as paradigmatic examples of such calculations.
A related approach, using the Nambu--Jona--Lasinio model,
can be exemplified by Refs.~\cite{cm:weise}.

   The successes of the potential model in describing heavy quarkonia are
well known. For light quarks, however, besides many important qualitative
successes, the potential model also exhibited weaknesses: first,
the uncertainty concerning the question
what potential can provide a realistic and yet reasonably tractable interaction
without many free parameters; and second, the noncovariance of
the instantaneous--potential approach to such highly relativistic constituents
as $u$, $d$, and $s$ quarks.

   As an example on the successful side, Le~Yaouanc et al. \cite{leyaouanc84}
demonstrated the appearance of dynamical (spontaneous) chiral--symmetry
breaking by generating the dynamical quark mass as well as the pion
as a Goldstone boson in the chiral limit. They studied
the power--law interaction ($V \propto r^\alpha$) between massless
quarks and antiquarks most exhaustively in the simplest case
of harmonic potential (V $\propto r^2$) where
the gap equation (Schwinger--Dyson equations) reduced from an integral equation
to a differential one. Adler
and Davis \cite{adler84} formulated the renormalization procedure
for the Coulomb--like potential and performed a concrete numerical calculation
for the linear confining quark--antiquark potential. In the latter case,
only infrared divergences appeared.
Trzupek \cite{trzupek89} applied their renormalization procedure to the
more realistic case of the funnel (linear--plus--Coulomb) potential where
ultraviolet (UV) divergences were present.
This situation was further complicated
if finite quark masses were present \cite{hirata87,alkofer88}. Nevertheless,
the aforementioned  weaknesses led to unsatisfactory quantitative results.
For example, the investigations of
Refs.~\cite{leyaouanc84,adler84,trzupek89} could
not yield the value for the
pion decay constant better than four to five times smaller than the
experimental one, when the meson spectrum was fitted correctly.
This was assumed to be the consequence of noncovariance \cite{leyaouanc84}.
(Furthermore, it was expected that finite current quark masses would improve
the results). Our attention was thus attracted by the covariant generalization
of the instantaneous--potential approach to bound states, formulated by
Pervushin and collaborators
\cite{kalinovsky89,pervushin90,kalinovsky90,kalinovsky91}.
This approach was first applied to the
harmonic potential as the simplest case \cite{amirkhanov90prep};
however the first
concrete and correct numerical results were obtained by our group
\cite{horvat91}. Besides covariance, the
effect of the finite current quark masses was also included \cite{horvat91},
while previous investigations in this line of research
\cite{leyaouanc84,adler84,trzupek89} had been concerned with massless
quarks only. Therefore, they had not been pertinent for studying the dependence
of the pion mass on the model parameters since in the chiral limit the pion
mass is vanishing. The mass of the pion from our Ref.~\cite{horvat91}
behaves as the square root of the current quark mass, which is precisely
the correct behavior of the (pseudo) Goldstone boson
(see, {\it e.g.}, \cite{gasser82}).
The pion decay constant, however, was found in Ref.~\cite{horvat91} to be
$F_\pi \approx 35~MeV$ for the harmonic--potential strength which reproduced
the experimental pion mass.
The result turned out to be practically the same as in 
Ref.~\cite{leyaouanc84}, $F_\pi=\sqrt{2}f_\pi=\sqrt{2}\,20~MeV\approx 28~MeV$.
The covariant approach removed certain ambiguities present in
the definition of the pion decay constant in Ref.~\cite{leyaouanc84},
but did not solve the problem of its too small value. Obviously,
further studies of the form of the interaction potential were needed. In the
present work we therefore examine the funnel (Coulomb--plus--linear) potential

\begin{equation}
V(r)=V_C(r)+V_L(r)=\frac{4}{3}(- \frac{\alpha_s}{r} + \sigma r)~,
\label{funnel}
\end{equation}

\noindent since it is known that, in QCD, the short--distance
interactions are dominated
by the Coulomb interaction, while in the long--distance (small momentum,
or $k\rightarrow 0$) regime, $\alpha_s$ times gluon propagator
seems to behave as $1/k^4$, corresponding to a linear confining potential
in the coordinate language. However, the ultraviolet
divergences caused by the Coulomb part pose some new difficulties.
In fact, the issue of renormalization of the bound--state equations
for quarkonium with instantaneous interaction is the main point
of this work. After sketching in Sec.~\ref{BILOCALS} how the representation
of mesons by bilocal fields leads to
the Schwinger--Dyson equation (SDE) and the Bethe--Salpeter equation (BSE)
in ladder approximation, in Sec.~\ref{SDE} we formulate a renormalization
scheme for the SDE with the funnel potential. We discuss the limitations
which such a scheme must suffer when various approximations are introduced.
We also compare our renormalization procedure with the ones used so far in this
context. In Sec.~\ref{BSE} the Salpeter equation for the pion is solved,
and in Sec.~\ref{PDC} the pion decay constant is obtained. In Sec.~\ref{PITOGG}
we calculate the $\pi^0\to\gamma\gamma$ decay width and conclude
in Sec.~\ref{CONCLUSION}.

\section{Mesons as bilocal fields}
\label{BILOCALS}

     When trying to model nonperturbative QCD, one may consider a
very general interaction kernel $K$ entering in the effective action:

\begin{eqnarray}
W_{eff}=\int d^{4}x \{ \bar{q}(x) [(i\rlap{$\partial$}/-\hat{m})-L(x)]q(x)
    \nonumber \\
+\frac{1}{2}\int d^{4}\! y \, \bar{q}_{\alpha_{2}}\!(y) q_{\beta_{1}}\!(x)
[K(x-y)]_{\alpha_{1},\beta_{1};\alpha_{2},\beta_{2}}
q_{\beta_{2}}\!(y) \bar{q}_{\alpha_{1}}\!(x) \}~.
\label{W_eff}
\end{eqnarray}

\noindent where $\hat{m}$ is the
current quark mass matrix, $\hat{m}=\mbox{\rm diag}(m_u, m_d, m_s)$.
$\alpha_i$ and $\beta_i$ are spinor indices, whereas color indices and
flavor indices are suppressed.
In (\ref{W_eff}) the summation over repeated indices is understood.
We assume that the interaction kernel $K(x-y)$ can lead to a bound
$q\bar{q}$ system. In (\ref{W_eff}) we have
introduced $L(x)$, an external operator coupled to the quark  current.
For example, it can be the leptonic current $l_\mu(x)$:

\begin{equation}
L(x)=\frac{G_{F}}{\sqrt{2}} l_{\mu} \gamma^{\mu} \frac{1-\gamma_{5}}{2}~,
\label{loc_op_w}
\end{equation}

\noindent or a photon $A_{\mu}(x)$:

\begin{equation}
L(x)=eA^{\mu}(x)\gamma_{\mu}=e\not\!\! A(x)~.
\label{loc_op_em}
\end{equation}

Such external operators will make possible the weak and radiative decays,
but the internal structure of hadronic bound states will be
dictated by the model kernel $K$.

One can construct a theory of meson bound states by eliminating bilinear
structures $q_{\alpha}(x) \bar{q}_{\beta}(y)$
in favor of bilocal fields $\chi_{\alpha,\beta}(x,y)$
\cite{kleinert76b,schrauner77,pervushin79,kugo90prep,aoki90},
(introduced through the path
integral in the generating functional)
and then integrating out the remaining quark fields. In this way
the action (\ref{W_eff}) becomes \cite{pervushin79}

\begin{eqnarray}
\lefteqn{W_{eff}[q,\bar{q}]\rightarrow W_{eff}[\chi]} \nonumber \\
& & = iN_c{\rm Tr}\ln[i\rlap{$\partial$}/-\hat{m}-L-\chi]
    + \frac{N_c}{2}(\chi,K^{-1}\chi)
 \nonumber \\
& & =N_c \{ i{\rm Tr}\ln(i\rlap{$\partial$}/-\hat{m})
    - i{\rm Tr}\sum^{\infty}_{n=1}
  {1\over n}
  [(i\rlap{$\partial$}/-\hat{m})^{-1}(L+\chi)]^n
  + \frac{1}{2} (\chi,K^{-1}\chi) \}~,
\label{W_eff_chi}
\end{eqnarray}

\noindent where we have suppressed all indices and used shorthand:

\begin{equation}
  (\chi,K^{-1}\chi)=
\int d^4x d^4y \chi_{\beta_1\alpha_2}(x,y)
K^{-1}_{\alpha_1\beta_1;\alpha_2,\beta_2}(x,y) \chi_{\beta_2\alpha_1}(y,x)~,
\end{equation}

\noindent and where Tr (with the capital ``T'') also includes the integration.
(Below, ``tr'', with small ``t'', will denote a trace not including
integration.)
We can drop the external (weak or electromagnetic) operator $L$ while
studying the bound--state equations determining the internal hadron
structure. We shall reinstate  $L$ later, while studying weak and
electromagnetic decays.

We determine the classical solution $\chi_{0}$ conveniently written as
$\chi_0(x-y) \newline \equiv \sum(x,y) - \hat{m} \delta^{4}(x-y)$, by
varying $W_{eff}$ with respect to $\chi$:

\begin{equation}
\frac{\delta W_{eff}[\chi]}{\delta \chi}=0~.
\label{dweff}
\end{equation}

	This yields the SDE for the
quark self--mass operator $\Sigma$ in the ladder approximation:

\begin{equation}
\Sigma(x-y)={\hat m}\delta^4(x-y)+iK(x-y){S}(x-y)~,
\label{SchDy}
\end{equation}

\noindent where the ``dressed'' quark propagator is defined by

\begin{eqnarray}
{S}^{-1}(x-y) & = & (i\rlap{$\partial$}/-\hat{m})\delta^{(4)}(x-y)
 -\chi_{0} \nonumber \\
 & = & i\rlap{$\partial$}/\delta^{(4)}(x-y)-\Sigma(x-y)~.
\label{mass_op_in}
\end{eqnarray}

   Next, we expand the fields $\chi$ in the action around the minimum,
$\chi(x,y) = \chi_{0}(x,y) + {\cal M}(x,y)
= \sum(x-y) - \hat{m}\delta^{(4)}(x-y) + {\cal M}(x-y)$.
As it will turn out that the fields ${\cal M}\!(x,y)$ represent mesons,
we separate the part of the action containing ${\cal M}\!(x,y)$:

\begin{equation}
 W_{eff}[\chi] = W_{eff}[\chi_0 + {\cal M}] = W_{eff}[\chi_0]
                 + {\widetilde W}_{eff}[{\cal M}]~,
\label{WtildeW}
\end{equation}

\noindent where this part is given by

\begin{equation}
{\widetilde W}_{eff}[{\cal M}] = \frac{N_c}{2}({\cal M},K^{-1}{\cal M})
 - i N_c \sum^{\infty}_{n=2} \frac{1}{n} {\rm Tr} \Phi^n
\label{tildeW}
\end{equation}

\noindent and where we have introduced

\begin{equation}
\Phi(x,y) \equiv \int d^4\!z {S}(x,z) {\cal M}(z,y)  \label{Phi}
\end{equation}

\noindent and, as before, ${\rm Tr} \Phi^n$ denotes

\begin{equation}
{\rm Tr} \Phi^n \equiv {\rm tr} \int d^4\!x_1\,d^4\!x_2 ... d^4\!x_n\,
   \Phi(x_1,x_2) \Phi(x_2,x_3) ... \Phi(x_n,x_1)~.
\label{TrPhin}
\end{equation}

Up to the terms of the order ${\cal O}({\cal M}^2)$, varying the action
$W_{eff}[\chi_0 + {\cal M}]$ or, equivalently, ${\widetilde W}_{eff}[\cal M]$,
with respect to
the fluctuations ${\cal M}(x,y)$ gives the BSE
in the ladder approximation for the bound state of a
quark and an antiquark whose spectra
and propagators are determined by Eq.~(\ref{SchDy}) and whose
flavors $a$, $b$ are now explicitly written out:

\begin{equation}
{\cal M}_{ab}(x,y)=iK(x-y)\int dx' dy'
  {S_a}(x-x')
  {\cal M}_{ab}(x',y'){S_b}(y'-y)~.
\label{m_xy}
\end{equation}

\noindent
Equation (\ref{m_xy}) is the BSE in
the somewhat  improved ladder approximation, since the quark propagators
in it are not the free, bare ones, but ${S}$, containing the nontrivial
self--energy function $\Sigma$.

   In this paper we work in the isosymmetric limit,
being concerned with pions only. Not having to distinguish quark
masses for different flavors allows us to simplify the notation in
the rest of this paper by dropping the indices $a$ and $b$.
(However, in processes
involving kaons, it will be necessary to keep track of quark
flavors carefully.)

\section{Covariant generalization \protect\\
	of the potential approach and the renormalized \protect\\
	Schwinger--Dyson equation for the funnel potential}
\label{SDE}

     To solve the SDE and the BSE in practice, one must limit
oneself to a tractable interaction kernel $K$, which is given by

\begin{equation}
K(k) = -i C_F g^2 \gamma^\mu \otimes \gamma^\nu D_{\mu\nu}(k)~.
\label{K_ladder}
\end{equation}

\noindent where $C_F$ is the second Casimir of the quark representation,
here $4/3$ for the case of SU(3) triplet,
$g$ is the strong coupling constant,
and $D$ is the gluon propagator. An instantaneous approximation to the
kernel $K$ leads to the potential model,

\begin{equation}
K(k) \approx
  i \gamma^0 \otimes \gamma^0 \widetilde{V}(\bbox{k})
- i  \gamma^j \otimes \gamma^l \widetilde{V}_T(\bbox{k}) \,
[ \delta^{jl} - \frac{k^j k^l}{|\bbox{k}|^2} ]~.
\label{K_ladder1}
\end{equation}

     Further approximation consists in neglecting
the transverse gluon exchange. Thus, the kernel $K$ becomes

\begin{equation}
K(k) \approx
  i \gamma^0 \otimes \gamma^0 \widetilde{V}(\bbox{k})~.
\label{K_ladder2}
\end{equation}

     Pervushin and his group
\cite{han87,pervushin90,kalinovsky90,kalinovsky91} have found that the
covariant generalization of the potential approach is possible, provided
that the kernel $K$ is of a special form $K^{n}$:

\begin{equation}
K^{n}(k) = i \rlap{${n}$}/ \otimes \rlap{${n}$}/
     \widetilde{V}(k_\perp)~,
\label{K_Pervushin}
\end{equation}

\noindent where ${n}$ is the timelike unit vector in the direction of the
total momentum of the bound system, ${n}^\mu=P^\mu/\sqrt{P^2}$.
For any vector $k^\mu=k_\parallel^\mu+k_\perp^\mu$,
the components parallel and perpendicular to this axis are

\begin{eqnarray}
\begin{array}{cc}
  k^{\mu}_{\parallel}={n}^{\mu} k_P,
& k_P=k\cdot{n} = k\cdot P/\sqrt{P^2},  \\
  k^{\mu}_{\perp}=k^{\mu}-k^{\mu}_{\parallel},
& k_{\perp}\cdot P=0~.
\end{array}
\label{kper_kpar}
\end{eqnarray}

	The relativistic covariant
formulation of potential models, given by the kernel (\ref{K_Pervushin}),
guarantees that the {\em correct dispersion relation}
for the momentum and mass of the bound state,
$P^2=M^2$, is fulfilled.

     For the interaction kernel of the form (\ref{K_Pervushin}),
the SDE~(\ref{SchDy})
Fourier--transformed to momentum space is

\begin{equation}
{S}^{-1}(p)=\rlap{$p$}/ - m +
i \int \frac{d^4k}{(2\pi)^4} \rlap{${n}$}/ {S}(k) \rlap{${n}$}/
\widetilde{V}(p_\perp -k_\perp)~.
\label{SD_k}
\end{equation}

     Equation (\ref{SD_k}) is valid for an arbitrary reference frame. For the
quarkonium rest frame, ${n}=(1,0,0,0)$, and in the chiral limit,
(\ref{SD_k}) reduces to the SDE
studied by, {\it e.g.}, Le~Yaouanc et al. \cite{leyaouanc84} in their
noncovariant approach with the power--law
interaction, $V(r)\propto r^\gamma$. In Ref.~\cite{horvat91} we have already
studied Eq.~(\ref{SD_k}) for the particularly
simple,  harmonic  case, with $\gamma=2$, where the integral SDE reduces to
differential equations.

     The UV divergences due to the Coulomb part
of the potential require renormalization and introduction of
counterterms. This will change Eq.~(\ref{SD_k}) and its
rest--frame version.
Following Ref.~\cite{adler84} and its generalization to the massive case
\cite{hirata87,alkofer88}, we use the equations for renormalized vector,
axial--vector and pseudoscalar vertices, and Ward identities, to set the
renormalized SDE in the ladder approximation,

\begin{equation}
{S}^{-1}(p) = Z_2 \rlap{$p$}/ - Z_m m
	  - i g^2 C_F \int \frac{d^4k}{(2\pi)^4}
	    \gamma^\mu {S}(k) \gamma^\nu D_{\mu\nu}(p-k)~,
\label{SD_k_ren}
\end{equation}

\noindent where $Z_2$ and $Z_m$ are the wave function and mass renormalization
constants defined by

\begin{eqnarray}
\begin{array}{ccc}
  S_{0} = Z_2 S &  \mbox{\rm and}
& m_{0} = \frac{\textstyle Z_m}{\textstyle Z_2} m~,
\end{array}
\label{S_and_m_renormalization}
\end{eqnarray}

\noindent where $S_{0}$ and $m_{0}$ are the bare quark propagator and
the bare quark mass, respectively. Neglecting the retardation effects,
Eq.~(\ref{K_ladder1}), and the transverse gluon exchange,
Eq.~(\ref{K_ladder2}), in (\ref{SD_k_ren})
we are provided with the renormalized version of
Eq.~(\ref{SD_k}):

\begin{equation}
{S}^{-1}(p) = Z_2 \rlap{$p$}/ - Z_m m
	  + i \int \frac{d^4k}{(2\pi)^4}
	    \gamma^0 {S}(k) \gamma^0 \widetilde{V}(\bbox{p}-\bbox{k})~.
\label{SD_k_ren_rest}
\end{equation}

	Since the perpendicular part of a four--vector
reduces to the corresponding three--vector in the quarkonium rest frame,
{\it e.g.}, $k_\perp \to \bbox{k}$, we will use the noncovariant notation
to the end of this section.
This will make easier the comparison
of our results with the results of other authors in this line of
research.
Of course, all the expressions can be generalized back to those
valid in moving frames, by substituting $\bbox{k}\to k_\perp$,
$k^0\to k_P$, and $\gamma^0\to \rlap{${n}$}/$.

     Let us demonstrate the multiplicative renormalizability (MR) of this
equation. The renormalization of the product $g^2 D$ is

\begin{equation}
g_{0}^2 D_{0} = (\frac{Z_1}{Z_2})^2 g^2 D~,
\end{equation}

\noindent where the subscript $0$ refers to the bare quantities
and $Z_1$ is the vertex renormalization constant. The
gauge invariance implies $Z_1=Z_2$ and the renormalization--group (RG)
invariance of $g^2 D$.
However, the ladder approximation is consistent with $Z_1=1$,
{\it i.e.}, with no vertex renormalization (see, {\it e.g.}, \cite{brown91}).
So, the renormalization of $g^2 D$, and hence of $\widetilde{V}$
should be

\begin{equation}
g_{0}^2 D_{0} = (\frac{1}{Z_2})^2 g^2 D~,
\end{equation}

\begin{equation}
\widetilde{V}_{0} = (\frac{1}{Z_2})^2 \widetilde{V}~.
\label{V_scal_law}
\end{equation}

	Now, suppose that $\{Z_2,Z_m\}$ and $\{Z_2^\prime,Z_m^\prime\}$
are two sets of renormalization constants. They may correspond to two
different renormalization scales $\mu$ and $\mu^\prime$.
Since there is a definite relationship between the bare and renormalized
quantities, we know the relationship between the quantities
renormalized by primed and unprimed $Z$'s.
Concretely, using (\ref{S_and_m_renormalization}) and (\ref{V_scal_law}),
we transform the SDE, Eq.~(\ref{SD_k_ren_rest}), and find that it does not
change its form, so that the MR holds. Before further discussion
of the MR, we shall rewrite
the SD equation (\ref{SD_k_ren}) using the conventional ansatz
of the quark propagator ${S}$ through the functions
$\omega$ and $\varphi$:

\begin{equation}
S^{-1}(k) = k^0 \gamma^0 -\omega(\bbox{k}) \zeta^{-2}(\bbox{k})~,
\label{QP_parametrization}
\end{equation}

\noindent where the matrix $\zeta$ is defined as

\begin{equation}
\zeta(\bbox{k}) =
	\sin\frac{1}{2}\varphi(\bbox{k}) -
	\hat{k}\cdot\bbox{\gamma} \cos\frac{1}{2}\varphi(\bbox{k})~.
\end{equation}

\noindent We can express the quark propagator as

\begin{equation}
S(k) = - \zeta(\bbox{k})
[
\frac{\frac{1}{2}(1+\gamma^0)} {\omega(\bbox{k}) - k^0 - i\varepsilon}
+
\frac{\frac{1}{2}(1-\gamma^0)} {\omega(\bbox{k}) + k^0 - i\varepsilon}
]
\zeta(\bbox{k})~.
\end{equation}

\noindent Inserting (\ref{QP_parametrization}) into the renormalized
SD equation (\ref{SD_k_ren}) yields the following integral equations
for $\omega$ and $\varphi$:

\begin{mathletters}
\label{SD_1_param}
\begin{eqnarray}
\omega(\bbox{p}) \sin\varphi(\bbox{p})
& - & Z_m m
+ \frac{1}{2} \int \frac{d^3\bbox{k}}{(2\pi)^3}
	\sin\varphi(\bbox{k}) \widetilde{V}(\bbox{p}-\bbox{k})=0~,
\label{SD_1_param(a)}
\\
\omega(\bbox{p}) \cos\varphi(\bbox{p})
& - & Z_2 |\bbox{p}|
+ \frac{1}{2} \int \frac{d^3\bbox{k}}{(2\pi)^3}
	\cos\varphi(\bbox{k}) (\hat{k}\cdot\hat{p})
	\widetilde{V}(\bbox{p}-\bbox{k})=0.
\label{SD_1_param(b)}
\end{eqnarray}
\end{mathletters}

     Additionally, the equation $(Z_2-1)p^0=0$ arises, {\it i.e.}, $Z_2=1$.
However, the Eq.~(\ref{SD_1_param(b)}) demands that $Z_2$ to be an infinite
constant. Adler and Davis \cite{adler84} have resolved that contradiction
by splitting $Z_2$ into two parts,

\begin{equation}
Z_2 \rlap{$p$}/ \to Z_0 p^0 \gamma^0 - Z \bbox{p}\cdot\bbox{\gamma}~,
\end{equation}

\noindent and setting $Z_0=1$. However, we have to remind ourselves
that $Z_2$ is the wave--function renormalization constant, which defines
the renormalization of the quark propagator, ${S}_{0} = Z_2 {S}$.
We are faced with a dilemma, namely, whether ${S}$ should be renormalized
with $Z_0$ or with $Z$. As one can expect, both possibilities change
the form of the renormalized SD equation (\ref{SD_k_ren_rest}).
Obviously, the consistency of MR is violated because one
has to split $Z_2$ into $Z_0$ and $Z$.
If unprimed and primed renormalization constants correspond to two
different renormalization scales $\mu$ and $\mu^\prime$, respectively,
this inconsistency automatically shows that the invariance with respect
to the changes of the renormalization scale $\mu$ is
lost, and we cannot use the renormalization group
(RG) equations to relate results for one arbitrary
scale $\mu$ to that for some other scale $\mu^\prime$.
Nevertheless, as noted by Brown and Dorey
\cite{brown91}, who explored
the consistency of the MR of the SDE when various approximations are made,
this does not mean that treatments that do fail such a consistency
test cannot be useful, merely that is more difficult to relate
their solutions to real physics.
Of course, we must investigate the scale dependence of our results.
We shall return to this point later in the text.

     A convenient renormalization prescription that determines
$Z$ and $Z_m$ uniquely is given, {\it e.g.},
in Refs.~\cite{georgi76,politzer76}.
The authors specify that the quark propagator ${S}(p)$, for a given
spacelike $p^2=-\mu^2$, agrees with free-field theory.
We adopt this choice, adjusted for a special form of the kernel
(\ref{K_Pervushin}) and the propagator (\ref{QP_parametrization}),

\begin{equation}
{S}^{-1}(p)|_{\textstyle |\bbox{p}|=\mu}=\rlap{$p$}/-m~.
\label{Politzer_rs}
\end{equation}

     Imposing (\ref{Politzer_rs}) on the SD equation (\ref{SD_k_ren})
yields the renormalization constants

\begin{mathletters}
\label{Ren_Const_1}
\begin{eqnarray}
Z= 1+\frac{1}{2\mu} \int \frac{d^3\bbox{k}}{(2\pi)^3}
	\cos\varphi(\bbox{k}) (\hat{\bbox{\mu}}\cdot\hat{\bbox{k}})
	\widetilde{V}(\bbox{\mu}-\bbox{k})~,
\label{Ren_Const_1(a)}
\\
Z_m= 1+\frac{1}{2m} \int \frac{d^3\bbox{k}}{(2\pi)^3}
	\sin\varphi(\bbox{k}) \widetilde{V}(\bbox{\mu}-\bbox{k})~.
\label{Ren_Const_1(b)}
\end{eqnarray}
\end{mathletters}

     Using (\ref{Ren_Const_1}), the SD equation
(\ref{SD_1_param}) becomes

\begin{mathletters}
\label{SD_1}
\begin{eqnarray}
\omega(\bbox{p})\sin\varphi(\bbox{p}) &-& m \nonumber \\
	&+& \frac{1}{2} \int \frac{d^3\bbox{k}}{(2\pi)^3}
	\sin\varphi(\bbox{k})
	[
	\widetilde{V}(\bbox{p}-\bbox{k}) -
	\widetilde{V}(\bbox{\mu}-\bbox{k})
	] = 0~,
\label{SD_1(a)}
\\
\omega(\bbox{p})\cos\varphi(\bbox{p}) &-& |\bbox{p}| \nonumber \\
	&+& \frac{1}{2} \int \frac{d^3\bbox{k}}{(2\pi)^3}
	\cos\varphi(\bbox{k})
	[
	(\hat{\bbox{p}}\cdot\hat{\bbox{k}}) \widetilde{V}(\bbox{p}-\bbox{k}) -
	\frac{|\bbox{p}|}{|\bbox{\mu}|}
		(\hat{\bbox{\mu}}\cdot\hat{\bbox{k}})
		\widetilde{V}(\bbox{\mu}-\bbox{k})
	] =0~.
\label{SD_1(b)}
\end{eqnarray}
\end{mathletters}

	For $m=0$, there is no mass renormalization, and the
functions $\varphi_D$ and $\omega_D$ of the dynamical quark propagator
${S}_D$ have to satisfy the equations

\begin{mathletters}
\label{SD_zm_1}
\begin{eqnarray}
\omega_D(\bbox{p})\sin\varphi_D(\bbox{p})
	& + & \frac{1}{2} \int \frac{d^3\bbox{k}}{(2\pi)^3}
	\sin\varphi_D(\bbox{k})
	\widetilde{V}(\bbox{p}-\bbox{k}) =0,
\label{SD_zm_1(a)}
\\
\omega_D(\bbox{p})\cos\varphi_D(\bbox{p}) & - & |\bbox{p}| \nonumber \\
	& + & \frac{1}{2} \int \frac{d^3\bbox{k}}{(2\pi)^3}
	\cos\varphi_D(\bbox{k})
	[
	(\hat{\bbox{p}}\cdot\hat{\bbox{k}}) \widetilde{V}(\bbox{p}-\bbox{k}) -
	\frac{|\bbox{p}|}{|\bbox{\mu}|}
		(\hat{\bbox{\mu}}\cdot\hat{\bbox{k}})
		\widetilde{V}(\bbox{\mu}-\bbox{k})
	] =0.
\label{SD_zm_1(b)}
\end{eqnarray}
\end{mathletters}

     Now, it is obvious that the SD equation (\ref{SD_1}) does not
reduce to Eq.~(\ref{SD_zm_1}) in the limit $m \to 0$,
because $\lim_{m \to 0} Z_m m \ne 0$. Moreover,
we see that $\varphi(\bbox{p})=\varphi_D(\bbox{p})$ and
$\omega(\bbox{p})=\omega_D(\bbox{p})$ is a solution to the
SD equation (\ref{SD_1}) for $m=m^\prime \equiv
\omega_D(\bbox{\mu})\sin\varphi_D(\bbox{\mu})$. So, the chiral limit
is reached in the SD equation (\ref{SD_1}) for $m \to m^\prime$
and not for $m \to 0$ as it should be. This is an artifact of the
renormalization prescription (\ref{Politzer_rs}). Pagels
\cite{pagels79a} showed that the normalization condition (\ref{Politzer_rs})
precluded the presence of a dynamically generated term
in the quark propagator ${S}(p)$,
and he argued in favor of Weinberg's zero--mass renormalization
scheme. However, it is possible to recover the proper chiral--limit
behavior by redefinition of the mass renormalization constant,
Eq.~(\ref{Ren_Const_1(b)}):

\begin{equation}
Z_m= 1 + \frac{1}{m}\omega_D(\bbox{\mu})\sin\varphi_D(\bbox{\mu}) +
	\frac{1}{2m} \int \frac{d^3\bbox{k}}{(2\pi)^3}
	\sin\varphi(\bbox{k}) \widetilde{V}(\bbox{\mu}-\bbox{k})~.
\label{Ren_Const_2}
\end{equation}

     This corresponds to the redefinition of the normalization
condition (\ref{Politzer_rs}):

\begin{equation}
{S}^{-1}(p)|_{\textstyle |\bbox{p}|=\mu}
	= [ \rlap{$p$}/-m
-\omega_D(\bbox{p})\sin\varphi_D(\bbox{p})]_{\textstyle |\bbox{p}|=\mu}~.
\label{Politzer_rs_mod}
\end{equation}

	Equation (\ref{SD_1(a)}) becomes

\begin{eqnarray}
  \omega(\bbox{p})\sin\varphi(\bbox{p})
&-& \omega_D(\bbox{\mu})\sin\varphi_D(\bbox{\mu})
- m \nonumber \\
	&+& \frac{1}{2} \int \frac{d^3\bbox{k}}{(2\pi)^3}
	\sin\varphi(\bbox{k})
	[
	\widetilde{V}(\bbox{p}-\bbox{k}) -
	\widetilde{V}(\bbox{\mu}-\bbox{k})
	] = 0~,
\label{SD_1_mod}
\end{eqnarray}

\noindent whereas Eq.~(\ref{SD_1(b)}) remains unchanged.

     The renormalization constants $Z$, Eq.~(\ref{Ren_Const_1(a)}),
and $Z_m$, Eq.~(\ref{Ren_Const_2}), are defined
in terms of the potential $\widetilde{V}$.
We are considering the case of funnel potential, which is
a sum of the Coulomb--like potential $\widetilde{V}_C$ and
the linear potential $\widetilde{V}_L$. The integrals involving the linear
potential are finite, so we can drop $\widetilde{V}_L$ from the definition of
the renormalization constants, {\it i.e.}, we can define

\begin{mathletters}
\label{Ren_Const_3}
\begin{eqnarray}
Z= 1+\frac{1}{2|\bbox{\mu}|} \int \frac{d^3\bbox{k}}{(2\pi)^3}
	\cos\varphi(\bbox{k}) (\hat{\bbox{\mu}}\cdot\hat{\bbox{k}})
	\widetilde{V}_C(\bbox{\mu}-\bbox{k})~,
\label{Ren_Const_3(a)}
\\
Z_m= 1
	-\frac{1}{2m} \int \frac{d^3\bbox{k}}{(2\pi)^3}
	\sin\varphi_D(\bbox{k}) \widetilde{V}_C(\bbox{\mu}-\bbox{k})
	+\frac{1}{2m} \int \frac{d^3\bbox{k}}{(2\pi)^3}
	\sin\varphi(\bbox{k}) \widetilde{V}_C(\bbox{\mu}-\bbox{k})~.
\label{Ren_Const_3(b)}
\end{eqnarray}
\end{mathletters}

     The first integral in the $Z_m$ definition, Eq.~(\ref{Ren_Const_3(a)}),
multiplied by $m$, now plays the role of $m^\prime$.
Omitting this term will cause the incorrect chiral limit
behavior we have met using the renormalization
constants (\ref{Ren_Const_1}). The renormalization constants
(\ref{Ren_Const_3}) are those we have actually used in our numerical
calculation. The corresponding SDE is

\begin{mathletters}
\label{SD_3}
\begin{eqnarray}
\omega(\bbox{p})\sin\varphi(\bbox{p}) &-&
	[
	m
	- \frac{1}{2} \int \frac{d^3\bbox{k}}{(2\pi)^3}
	\sin\varphi_D(\bbox{k})
	\widetilde{V}_C(\bbox{p}-\bbox{k})
	] \nonumber \\
	&+& \frac{1}{2} \int \frac{d^3\bbox{k}}{(2\pi)^3}
	\sin\varphi(\bbox{k})
	[
	\widetilde{V}_C(\bbox{p}-\bbox{k}) -
	\widetilde{V}_C(\bbox{\mu}-\bbox{k})
	] \nonumber \\
	&+& \frac{1}{2} \int \frac{d^3\bbox{k}}{(2\pi)^3}
	\sin\varphi(\bbox{k})
	\widetilde{V}_L(\bbox{p}-\bbox{k}) = 0~,
\label{SD_3(a)}
\\
\omega(\bbox{p})\cos\varphi(\bbox{p}) &-& |\bbox{p}| \nonumber \\
	&+& \frac{1}{2} \int \frac{d^3\bbox{k}}{(2\pi)^3}
	\cos\varphi(\bbox{k})
	[
	(\hat{\bbox{p}}\cdot\hat{\bbox{k}}) \widetilde{V}_C(\bbox{p}-\bbox{k}) -
	\frac{|\bbox{p}|}{|\bbox{\mu}|}
		(\hat{\bbox{\mu}}\cdot\hat{\bbox{k}})
		\widetilde{V}_C(\bbox{\mu}-\bbox{k})
	] \nonumber \\
	&+& \frac{1}{2} \int \frac{d^3\bbox{k}}{(2\pi)^3}
	\cos\varphi(\bbox{k})
	(\hat{\bbox{p}}\cdot\hat{\bbox{k}})
	\widetilde{V}_L(\bbox{p}-\bbox{k}) =0~.
\label{SD_3(b)}
\end{eqnarray}
\end{mathletters}

	Let us now make a comparison with the renormalization
prescription of Refs.~{\cite{trzupek89,hirata87,kalinovsky92prep}.
The renormalization constant $Z$ used by them is

\begin{equation}
Z= 1 + \frac{1}{2|\bbox{p}|} \int \frac{d^3\bbox{k}}{(2\pi)^3}
     (\hat{p} \cdot \hat{k})
     \widetilde{V}_C(\bbox{p}-\bbox{k})~.
\label{Z_FiMa}
\end{equation}

	One way in which Eq.~(\ref{Z_FiMa}) differs from our $Z$,
Eq.~(\ref{Ren_Const_3(b)}), is the absence of the factor
$\cos\varphi(\bbox{k})$. This is not a critical difference,
as the modification of Eq.~(\ref{Ren_Const_3}) by making
the substitution $\cos\varphi(\bbox{k}) \to 1$
would change $Z$ only by a finite value.
As noted by Adler and Davis \cite{adler84},
the crucial point is
that Eq.~(\ref{Z_FiMa}) corresponded to a momentum--dependent subtraction
of UV divergences when the coupling
constant $\alpha_s$ was running, {\it i.e.}, when $\alpha_s$ was
momentum dependent. (For that reason, they
rejected the SDE with the Coulomb--like interaction
of Finger--Mandula \cite{finger82}.) For $\alpha_s=\mbox{\rm const}$,
however, they noted that the infinite part of (\ref{Z_FiMa}) defined a
momentum--independent UV subtraction.  Nevertheless, this counterterm
still has a momentum--dependent finite part. This is conceptually
objectionable even if the UV infinity is successfully subtracted.

     The other possibility considered by
Adler and Davis \cite{adler84} for the Coulomb--like
counterterm, again in connection with the unrenormalized equations
of Amer {\it et al.} \cite{amer83,amer83b}. This alternative
$Z$ is given by Eq.~(2.17) of Adler and Davis \cite{adler84} and is
obviously a completely momentum--independent
choice. It can be shown that it is given
by (\ref{Z_FiMa}) when $|\bbox{p}|$ tends to zero:

\begin{equation}
Z= 1 + \lim_{|\bbox{p}|\to 0}
\frac{1}{2|\bbox{p}|} \int \frac{d^3\bbox{k}}{(2\pi)^3}
     (\hat{\bbox{p}} \cdot \hat{\bbox{k}})
     \widetilde{V}_C(\bbox{p}-\bbox{k})~.
\label{Z_AdDe}
\end{equation}

     It is evident that (\ref{Z_AdDe}) is momentum independent simply
because $|\bbox{p}|$ is fixed to a specific value -- zero.
However, there is a subtlety that seems to have been unnoticed
so far. After performing angular integration in Eq.~(\ref{Z_AdDe}),
we obtain

\begin{equation}
Z = 1 - \frac{4\alpha_s}{3\pi} \int_0^\infty \frac{dk}{k}~.
\label{Z_IR_sing}
\end{equation}

\noindent This means that, for $k\rightarrow 0$, the UV renormalization
constant (\ref{Z_AdDe}) introduces a new IR divergence,
which has not been present in the original SDE! If $\alpha_s$ is not
constant but runs with $k$, then $\alpha_s$ have to be under the integral;
however, this does not change the situation.
One possibility for avoiding troubles for $p \rightarrow 0$ is
to introduce an IR cutoff, but this requires a new unphysical parameter.
Alkofer and Amundsen \cite{alkofer87} used
the renormalization constant (\ref{Z_AdDe}) but
no IR cutoff, which makes their
results suspicious
   \footnote{Alkofer and Amundsen studied the temperature-- and
   momentum--dependent $\alpha_s$, but the claim of the singularity
   in $Z$ for $p\rightarrow 0$ still holds.}.
On the other hand,
our renormalization constant (\ref{Ren_Const_3}) also
reduces to (\ref{Z_AdDe}) as $|\bbox{\mu}| \to 0$ apart
from the factor $\cos\varphi(\bbox{k})$ in the integrand.
For $|\bbox{k}| \to 0$, $\varphi(\bbox{k}) \to \pi /2$ and
$\cos\varphi(\bbox{k}) \to 0$,
which makes the integration in (\ref{Ren_Const_3}) IR regular.

   To summarize, our $Z$ differs from that used (so far) most frequently
in this line of research \cite{adler84,hirata87,trzupek89,kalinovsky92prep},
in that the variable momentum is replaced by the fixed renormalization
scale $\mu$. At the same time it encompasses the alternative
possibility \cite{adler84,alkofer87} for $Z$ in the limit $\mu \to 0$.
The finite part is chosen so that no new IR divergence arises in this limit.

     Our solutions are given in Fig.~\ref{fig:solutions},
while numerical methods are detailed in the Appendix.

     The linear potential $\widetilde{V}_L$ makes the integrals
in the SDE (\ref{SD_3}) IR divergent. The formulae
(\ref{SD_3}) can be rewritten such that the equations for $\varphi$ and
$\omega$ are separated. The equation for $\varphi$ needs no IR regularization
\cite{adler84}. The other equation expresses $\omega$ as a functional of
$\varphi$ and requires IR regularization. We adopt the procedure used  by
Le~Yaouanc et al. \cite{leyaouanc84}, where

\begin{equation}
V_L(r)=\frac{4}{3} \sigma r = \frac{4}{3}
\sigma \lim_{\xi\rightarrow 0} \frac{2}{\xi^2} \,
	\frac{e^{-\xi r} -1 +\xi r}{r}~,
\label{r_ley}
\end{equation}

\noindent which gives the IR--finite quark energy $\omega(k)$.
For light quarks, $\omega(k)$ is negative for small momenta,
as expected from earlier works, {\it e.g.}, \cite{leyaouanc84,amirkhanov90prep}.
This is in fact a signature of
dynamical chiral--symmetry breaking.

     To conclude this section, we have successfully solved
the SDE for massive quarks with the funnel--potential interaction,
treating IR and UV divergences carefully. With the solution
for $\varphi(p)$, we can proceed to solving and examining
the Salpeter equation for a light pseudoscalar quark--antiquark bound state.

\section{The Salpeter equation for the pion}
\nopagebreak
\label{BSE}

   Solving the SDE (\ref{SD_3}) for $\varphi(k)$ provided us with
the necessary input for solving the BSE~(\ref{m_xy}).
By Fourier transforming to momentum space, this equation becomes

\begin{equation}
\Gamma(q|P)=i\int \frac{d^4q^\prime}{(2\pi)^4}K(q-q^\prime)
{S}(q^\prime+\frac{P}{2})
\Gamma(q^\prime|P)
{S}(q^\prime-\frac{P}{2})~,
\label{gamma_ab}
\end{equation}

\noindent where $2q=p_a-p_b$ and $P=p_a+p_b$ are the relative and the total
momentum of the quark--antiquark pair, respectively, and
$\Gamma(q|P)$ is the momentum
vertex function of their bound state.
Using the instantaneous potential for the
interaction kernel reduces the BSE to the Salpeter equation. This
equation is still manifestly Lorentz covariant for the special form
(\ref{K_Pervushin}) of the instantaneous interaction, as shown by
Refs.~\cite{pervushin90,kalinovsky90,kalinovsky91}.
It is convenient to write this equation in terms of the quarkonium
wave function $\Psi(q_\perp)$ or its transformed mate $\psi^P$:

\begin{eqnarray}
\Psi^P(q_\perp) & = &
i\int\frac{dq_P}{(2\pi)} [ {S}(q+\frac{P}{2})
\Gamma(q_\perp|P){S}(q-\frac{P}{2})] \label{Psi} \\
& \equiv & \zeta(q_\perp)\psi^{P}(q_\perp)\zeta(q_\perp)~.
\nonumber
\end{eqnarray}

   In this work we are interested only in pseudoscalar mesons.
It can be shown that
$\mbox{$\not \!\! P$} \psi^P=-\psi^P \mbox{$\not \!\! P$}$, so that
the decomposition of $\psi^P$ for pseudoscalar mesons
in the Dirac matrices is simply

\begin{equation}
\psi^P(q_\perp)=\gamma_5
[L_1(q_\perp)+\frac{\mbox{$\not \!\! P$}}{\sqrt{P^2}}L_2(q_\perp)]~.
\label{psiL1L2}
\end{equation}

   The BSE~(\ref{gamma_ab}) written in terms of (\ref{Psi})
and (\ref{psiL1L2}) and for the form
(\ref{K_Pervushin}) is boost invariant. To solve it
for $L_1$ and $L_2$, we
are free to choose the rest frame of the bound system
$q_\perp=(0,\bbox{q})$, $P=(M_\pi,\bbox{0})$.
The mass $M_\pi$ of the pseudoscalar bound state, the pion, is the eigenvalue
of the BSE. (The treatment of the kaon is in principle identical to
that of the pion, except that one of the quarks would have to be
the significantly heavier strange quark).

	Inserting (\ref{Psi}), (\ref{psiL1L2}),
and (\ref{K_Pervushin}) in (\ref{gamma_ab})
yields the coupled system of integral equations, which at
first sight seems to be IR infinite,

\begin{mathletters}
\label{L1L2}
\begin{eqnarray}
M_\pi L_2(p_\perp) + 2\omega(p_\perp) L_1(p_\perp)
	&+& \int\frac{d^3 k_\perp}{(2\pi)^3} \widetilde{V}(p_\perp-k_\perp)
	L_1(k_\perp)=0,
\label{L1L2(a)}
\\
M_\pi L_1(p_\perp) + 2\omega(p_\perp) L_2(p_\perp)
	&+& \int\frac{d^3 k_\perp}{(2\pi)^3} \widetilde{V}(p_\perp-k_\perp)
	[
	\sin\varphi(p_\perp) \sin\varphi(k_\perp) \nonumber \\
	&-& (\hat{p}_\perp \cdot \hat{k}_\perp)
	\cos\varphi(p_\perp) \cos\varphi(k_\perp)
	] L_2(k_\perp)=0.
\label{L1L2(b)}
\end{eqnarray}
\end{mathletters}

	However, as in the SDE, the IR
infinity of the unregularized quark energy $\omega(k)$ cancels the IR
infinity of the integrals with kernels (see \cite{adler84}).

   The numerics for solving (\ref{L1L2}) is discussed
in the Appendix along with the numerics for the SDE. The solutions
$L_1$ and $L_2$ are displayed in Fig.~\ref{fig:solutions}.

	Having solved the massless version of the SDE,
Eq.~(\ref{SD_3}), we have found that
$\omega_D(\mu)\sin\varphi_D(\mu)\to 0$ as $\mu\to \infty$. So, in the limit of
infinitely large renormalization point $\mu$, the normalization condition
(\ref{Politzer_rs}) approaches the normalization condition
(\ref{Politzer_rs_mod}).
The improper behavior of the theory, artificially generated by
the normalization
condition (\ref{Politzer_rs}), will disappear for $\mu\to \infty$.
Fig.~\ref{fig:mp1mquark} shows the pion mass $M_\pi$ as a function
of the renormalized quark mass $m$ for three different renormalization
prescriptions. The dotted lines relate to the normalization condition
(\ref{Politzer_rs}), slightly modified by leaving out the linear potential
from definition of the renormalization constants,
and $\mu=1~GeV$. In this case, $M_\pi\to 0$ for
$m\to m^\prime(\mu) = \omega_D(\mu)\sin\varphi_D(\mu)$
($=0.70~MeV$ for $\alpha_s=0.4$ and $4.2~MeV$ for $\alpha_s=0.8$).
The dashed lines relate to the
same normalization condition (\ref{Politzer_rs}), but for $\mu=5~GeV$.
Now, $\omega_D(\mu)\sin\varphi_D(\mu) \sim 0.1~MeV$ and the improper
behavior of $M_\pi(m)$ has almost disappeared. However, if we use
the normalization
condition (\ref{Politzer_rs_mod}), we obtain $\lim_{m\to 0} M_\pi(m)=0$
for arbitrary choice of $\mu$
(solid lines in Fig.~\ref{fig:mp1mquark}).
This particular scheme is therefore preferred, and
we will discuss our results for the pion decay
constant $F_\pi$ and the $\pi^0\to\gamma\gamma$ decay width using
this scheme.

   Finally, let us remark that from the low--momentum behavior of the
SD solution $\varphi$ one can read off the constituent quark mass $m^\star$
as defined by Adler and Davis \cite{adler84}. Reference \cite{adler84} implies
$m^\star=-1/\varphi^\prime(0)$. $m^\star$ is a linear function of $m$ and is
depicted in Fig.~\ref{fig:mc1mquark}. These results are consistent with the
constituent mass of Ref.~\cite{adler84} which is $m^\star=70~MeV$ for
the massless case and the pure linear potential with $\sigma=(350~MeV)^2$.
As a function of $\sigma$,
$m^\star$ grows as square root (Fig.~\ref{fig:mc1sigma}).

\section{The pion decay constant $\uppercase{F}_\pi$}
\label{PDC}

     So far we have been concerned with describing the hadronic structure
which is determined in this context by the quark--quark
interaction $K$. Therefore, for simplicity, we have so far
omitted the external local operator $L(x)$, which is used in the present
approach to describe weak and radiative decays. However, rederiving the
bilocal action $\widetilde{W}_{eff}$ in the presence of $L(x)$ shows that
we can consistently reinstate $L(x)$ by the substitution

\begin{equation}
{\cal M}(x,y) \rightarrow {\cal M}(x,y) + L(x)\delta^{(4)}(x-y)~.
\end{equation}

\noindent The matrix element for the leptonic decay of
pseudoscalar mesons is then \cite{horvat91}

\begin{eqnarray}
<l^\pm\nu_l|\widetilde{W}_{eff}|\pi^\pm> & = &
<l^\pm\nu_l|
 \frac{i}{2}N_c\,{\rm Tr}({S}({\cal M}+L))^2
           |\pi^\pm> \\ & = &
<l^\pm\nu_l|
 iN_c\,{\rm Tr}({S}{\cal M}{S} L)
           |\pi^\pm>~.   \label{M.E.}
\end{eqnarray}

\noindent It is expressed through the axial--current matrix element which is
conveniently parametrized by the pion decay constant $F_\pi$.
Evaluating  (\ref{M.E.})
thus yields \cite{pervushin90} for equal $u$ and $d$ quark masses

\begin{equation}
F_\pi = \frac{4N_c}{M_\pi}\int \frac{d^3q_\perp}{(2\pi)^3}L_2(q_\perp)
\sin\varphi(q_\perp)~.
\label{Fpi}
\end{equation}

\noindent In Ref.~\cite{horvat91} we obtained the first correct numerical
results for the decay
constants in the bilocal approach with the harmonic potential not only for the
pion, but also for the kaon and their radial excitations.
(Of course, (\ref{Fpi})
was generalized for different quark masses because of the kaon.)

   The variation of $F_\pi$ with $\sigma$ (for $m=7~MeV$) is given in
Fig.~\ref{fig:fp1mquark}. $F_\pi$ grows monotonically with $\sigma$
roughly like a square root, Fig.~\ref{fig:fp1sigma}.

   At $\mu=1~GeV$ and for fixed $\sigma=(350~MeV)^2$, our $M_\pi$ is
fitted to the experimental value for $m=1.92~MeV$ when $\alpha_s=0.4$, and
for $m=2.80~MeV$ when $\alpha_s=0.8$.
Unfortunately,
Fig.~\ref{fig:fp1mquark} shows that $F_\pi$ is then too small,
being typically about $20-30~MeV$.
We remark that in the earlier treatments of the linear or funnel
potential $F_\pi$ has been even smaller, {\it e.g.},
$F_\pi=\sqrt{2}f_\pi=16~MeV$ in Ref.~\cite{adler84}, and
$F_\pi=(16-34)~MeV$ in Ref.~\cite{trzupek89}.
However, very close to the massless regime,
at $m=0.22~MeV$
for $\alpha_s=0.4$ and at $m=0.41~MeV$ for $\alpha_s=0.8$, the correct
experimental ratio $M_\pi/F_\pi$ is obtained. This means that we can fit
both $M_\pi$ and $F_\pi$ provided we rescale all dimensional quantities,
including $\mu$. Therefore, if we rescale by a factor of 7.9, we get
$M_\pi=140~MeV$ and $F_\pi=132~MeV$ at $m=1.71~MeV$,
$\sigma=(7.9\times 350~MeV)^2$ at the renormalization scale $\mu=7.9~GeV$.
Similarly, for $\alpha_s=0.8$ we
again practically reproduce the experimental values if we increase the
scale by a factor of $6$: at $\mu=6.0~GeV$, $m=2.47~MeV$,
and $\sigma=(6.0\times 350~MeV)^2$,
we get $M_\pi=140~MeV$ and $F_\pi=132~MeV$. Clearly, continually varying
$\alpha_s$ and $\sigma$ would give experimental values of $M_\pi$ and
$F_\pi$ for
continuum of different quark masses $m$ but also different scales $\mu$.

\section{The $\pi^0 \rightarrow\gamma\gamma$ decay width}
\label{PITOGG}

   If the external operator is taken to be $L(x)=QA_\mu(x)\gamma^\mu$,
where $Q=e\,\mbox{\rm diag}(Q_u,Q_d,Q_s)=e\,\mbox{\rm diag}(2/3,-1/3,-1/3)$,
it will make possible the
radiative processes as the $\pi^0 \rightarrow\gamma\gamma$ decay
computed in \cite{horvat91} for the
harmonic interaction. We consider the present case of the funnel potential
more realistic but still extremely oversimplified, so that we do not intend
that our calculation should compete with the standard description
via the Adler--Bell--Jackiw anomaly and PCAC, which yields an almost
experimental decay width. We calculate this decay more as a further test of
the quality of the funnel interquark interaction.
We should however stress that our approach is in a way more ambitious
than most of the other approaches, including the standard anomaly calculation:
these, namely, always contain the
step when one actually parametrizes the unknown hadronic structure with the
pion decay constant $F_\pi$. On the contrary, the present calculation is not
parametrizing but trying to describe the pion structure and in this
respect it is more microscopic.

   The $\pi^0 \rightarrow\gamma\gamma$ transition is caused
by the cubic term from $\widetilde{W}_{eff}$
because it contains subterms with one meson
bilocal ${\cal M}$ and two photon fields $A_\mu$. The transition
matrix element is thus

\begin{equation}
{\cal A}_{\pi^0\gamma\gamma}=<\gamma(k,\sigma)\gamma(k^\prime,\sigma^\prime)|
   iN_c\mbox{\rm Tr}\left[ {\cal M} {S} Q \mbox{$\not \!\! A$}
              {S} Q \mbox{$\not \!\! A$} {S} \right]
   |\pi^0(P)>~,
\label{ampl}
\end{equation}

\noindent where the symbol ``Tr'' also includes the integrations
over coordinates. Equation (\ref{ampl}) in fact corresponds to the $\gamma_5$
triangle graph except that the propagator lines emanate out
of a pseudoscalar bilocal bound--state vertex
and that these propagators are not free but dressed ones.
Transforming to momentum space,

\begin{equation}
{\cal A}_{\pi^0\gamma\gamma}=\frac{(2\pi)^4\delta^{(4)}(P+k+k')}
   {\sqrt{(2\pi)^32^3P_0k_0k_0'}} \,\, {\cal T}_{\pi^0\gamma\gamma}~,
\end{equation}

\begin{equation}
{\cal T}_{\pi^0\gamma\gamma}\equiv
      2iN_ce^2\frac{Q^2_u-Q^2_d}{\sqrt{2}}
\epsilon_\mu(k,\sigma)\epsilon_\nu(k',\sigma')I^{\mu\nu}~,
\end{equation}

\noindent where

\begin{equation}
I^{\mu\nu}
\equiv \int \frac{d^4q}{(2\pi)^4} {\rm tr}
     [
     \Gamma(q_\perp|P)
     {S}(q-P)
     \gamma^\mu
     {S}(q+k')
     \gamma^\nu
     {S}(q)
     ] \,\,\, ~.
\end{equation}

Inserting $\Gamma$ and ${S}$,
rearranging, and integrating over the parallel component $q_P$ and
performing the spinor trace,

\begin{equation}
I^{\mu\nu}=4\epsilon^{\alpha\beta\mu\nu}\frac{P_\alpha}{M_\pi}I_\beta~,
\end{equation}

\begin{equation}
\Re {\rm e}\, I_\beta=\int\frac{d^3q_\perp}{(2\pi)^3}
{\cal J}_\beta(q_\perp,k_\perp',\varphi)
	\frac{
L_2(q_\perp)[E(q_\perp)+E((q+k')_\perp)]
	-L_1(q_\perp)\frac{\textstyle{M_\pi}}{\textstyle{2}}
	     }
	     {
[E(q_\perp)+E((q+k')_\perp)]^2-\frac{\textstyle{M^2_\pi}}{\textstyle{4}}
	     }~,
\end{equation}

\begin{displaymath}
\Im {\rm m}\, I_\beta= -\frac{\pi}{2}\int\frac{d^3q_\perp}{(2\pi)^3}
{\cal J}_\beta(q_\perp,k_\perp',\varphi)
\{  \delta[  E(q_\perp)+E((q+k')_\perp)+\frac{M_\pi}{2}  ]
\end{displaymath}
\begin{equation}
    \times[L_1(q_\perp)+L_2(q_\perp)]
  -\delta[  E(q_\perp)+E((q+k')_\perp)-\frac{M_\pi}{2}  ]
    [L_1(q_\perp)-L_2(q_\perp)] \}~,
\label{ImI}
\end{equation}

\begin{eqnarray}
{\cal J}_\beta(q_\perp,k_\perp',\varphi)=
-\frac {[(q+k')_\perp)]_\beta} {|(q+k')_\perp)|}
\sin\varphi(q_\perp)\cos\varphi((q+k')_\perp)  \nonumber \\
+\frac{(q_\perp)_\beta}{|q_\perp|}
\sin\varphi((q+k')_\perp)\cos\varphi(q_\perp)~.
\end{eqnarray}

\noindent Since $I_\beta$ is a function of $M_\pi$ and of only one four vector,
($k_\perp^\prime$),

\begin{equation}
   I_\beta=(k'_\perp)_\beta\,{\cal C}[E,L_1,L_2,\varphi,M_\pi]~,
\label{I_beta}
\end{equation}

\noindent where ${\cal C}$ is a dimensionless Lorentz scalar functional of
$E$, $L_1$, $L_2$, and $\varphi$ and a
function of $M_\pi$  and $k_\perp^\prime$. We can thus extract it
numerically by evaluating
(\ref{I_beta}) in, say, the rest frame as the easiest one. Noting that
$\epsilon^{\alpha\beta\mu\nu}P_\alpha(k'_\perp)_\beta=
- \epsilon^{\alpha\beta\mu\nu}k_\alpha k'_\beta$,
summing $|{\cal A}_{\pi^0\gamma\gamma}|^2$ over the  polarizations
$\sigma$, $\sigma^\prime$, and
integrating over the phase space of the outgoing photons yields

\begin{equation}
  \Gamma(\pi^0\rightarrow\gamma\gamma) = \alpha^2 M_\pi 8 \pi |{\cal C}|^2~.
\label{Gpigg}
\end{equation}

   The dependence of $\Gamma$ on $m$ (for fixed $\mu=1~GeV$ and
$\sigma=(350~MeV)^2)$ is depicted in Fig.~\ref{fig:pigg1mquark}.
The experimental value
$\Gamma_{exp}=(7.7\pm 0.5)~eV$ is reached for $\alpha_s=0.4$ at
$m=11.7~MeV$ and for $\alpha_s=0.8$ at $m=12.1~MeV$. For such
quark masses, $M_\pi$ is already too large, while $F_\pi$ is still
too small. Since $\Gamma$ falls rather quickly with $m$,
in a good approximation proportional
to $m^{3/2}$, it is far too small in the range of $m^\prime$s for
which we could fit both $M_\pi$ and $F_\pi$
using rescaling, as shown in the preceding section.

   On the other hand, for $\mu=6.7~GeV$, $m=25.6~MeV$, and $\alpha_s=0.4$
(or $\mu=5.0~GeV$, $m=23.3~MeV$, and $\alpha_s=0.8$)
we can also fit $F_\pi$ and $\Gamma$ to experiment
using rescaling, but then $M_\pi$ becomes too large by a factor of $7$ to $10$.
The variation of $\sigma$ with fixed $m=7~MeV$ does not
yield better results either. This is reminiscent of the situation in the
harmonic--oscillator case \cite{horvat91} where it was impossible to
fit $M_\pi$, $F_\pi$ and $\Gamma$ simultaneously. In this case, however,
we have not explored the whole parameter space and it is still
possible that appropriate choice of $\alpha_s$, $m$, $\sigma$, and
$\mu$ would fit all three quantities simultaneously.

   The fact that $\Gamma$ depends on the current quark mass as
$m^{3/2}$, is another manifestation of consistency with
PCAC, as discussed in Sec.~V of Ref.~\cite{horvat91}.

\section{Conclusion}
\label{CONCLUSION}

   We have studied the pion as the quarkonium bound by the funnel potential.
We did it using the covariant generalization of the instantaneous--potential
model proposed in the framework of the effective bilocal Lagrangian
\cite{pervushin90,kalinovsky90,kalinovsky91,amirkhanov90prep}.
Provided that the covariant BSE solutions thus obtained describe
bound states sufficiently well, they can be useful
for understanding what happens
in the experiments probing the quark substructure. In these experiments
both nonperturbative
bound--state effects and relativistic recoil effects can be important, as in
the program of CEBAF, for example.

   Our work has extended the line of research which employs instantaneous
interquark potentials
({\it e.g.}, Refs.~\cite{leyaouanc84,adler84,trzupek89,hirata87,alkofer88}
and \cite{alkofer87})
by using the boost--invariant potential ansatz,
by introducing and analyzing the effects of the finite quark mass,
and by improving and generalizing the subtraction procedure for
UV divergences caused by the Coulomb part of the interaction.
The initial choice for the renormalization prescription,
essentially following from the standard prescription where infinities are
subtracted at a spacelike point \cite{georgi76}, has been shown to preclude the
correct chiral limit. However, by appropriately modifying the renormalization
prescription, we have been able to recover the correct chiral limit. With
this (naturally preferred) scheme, our pion exhibits the qualitatively
correct (pseudo)--Goldstone behavior. We have also shown how the MP
breaking occurs in this context and explored the dependence of the
results on the subtraction point $\mu$. In particular, we have found that,
as $\mu$ grows, the results in the first renormalization scheme
gradually approach the results in the second, preferred scheme.

   As far as the quantitative results are concerned, the description of
the pion is still not satisfactory, as the usage of the funnel potential
has not (yet) resulted in the improvement of $M_\pi$, $F_\pi$, and
$\Gamma(\pi^0\rightarrow\gamma\gamma)$ with respect to the values we obtained
with the harmonic interaction \cite{horvat91}.

   These results are not definitive as we have not yet systematically
explored the parameter space. This may be one task
for the future work, but further search for an improved form of the potential
certainly remains as the other, and even more important task.

\acknowledgments

   This work is partially supported by the
EC contract CI1*--CT91--0893 (HSMU). D. Klabu\v{c}ar also
acknowledges the partial support of the NSF under the contract JF 899-31.

\appendix
\section*{}

   The SD equation (\ref{SD_3}) and the BS equation
(\ref{L1L2}) are nonlinear integral equations with singular
kernels. In order to
solve these equations numerically, we discretize them, {\it i.e.}, we
approximate integral equations with a finite set of coupled nonlinear
equations. This is realized by discretization of the momentum variable,

\begin{eqnarray}
k_i=k_{max}(\frac{i-1}{N})^n & \,\,\,\,\,\,\, (i=1,...,N+1)~.
\label{mesh}
\end{eqnarray}

   $k_{max}$ is the ``numerical'' cutoff;
it should be chosen large enough to eliminate the effects of
the boundary condition at ``infinity''. $N+1$ is the number of points.
For $n=1$ Eq.~(\ref{mesh}) corresponds to an equidistant mesh, {\it i.e.},
uniformly distributed points, while for $n>1$ the points are denser near
$k=0$. A finite set of variables is defined, {\it e.g.}, for the SDE we define
$\varphi_i=\varphi(k_i)$, $i=1,...,N+1$. The values $\varphi_1$
and $\varphi_{N+1}$ are
fixed by the boundary conditions $\varphi(0)=\pi/2$ and
$\lim_{k\rightarrow +\infty}\varphi(k)=0$ (see, {\it e.g.}, \cite{leyaouanc84}).
Equation~(\ref{SD_3}) must be satisfied for every $k_i$, $i=2,...,N$;
thus we have obtained a system of $N-1$ nonlinear coupled equations for
$\varphi_1,...,\varphi_{N-1}$.

   A problem arises from the singular nature of the integration kernel.
The integrals have to be calculated as principal values. So, we define
the continuous function $\varphi(k)$ as a cubic spline of
$\varphi_1,...,\varphi_{N+1}$ and perform integration using
Gaussian quadrature, adapted for principal--value calculation.
The system of $N-1$ coupled nonlinear equations is solved using
a modification of Brent's methods.
A similar procedure was applied to the BSE (\ref{L1L2}), treating
$M_\pi$ as an additional variable.

   For $k_{max}$ larger than a few $MeV$, the solutions has been found to be
independent of $k_{max}$. For an equidistant mesh, $n=1$, the number of points
have to be $N\ge 100$ to obtain a solution independent of $N$. For a mesh
with $n=3$, this is $N\approx 25$. The meshes with larger $N$
have an inadequate
distribution of points in respect to the solution $\varphi(k)$. This
optimization is important because the computer time consumed
behaves as $\propto N^2$.

%\bibliography{funnel,comment}

\begin{figure}
\caption{Solutions $\varphi$ of the SDE, Eq.~(\protect\ref{SD_3}),
and $L_1$ and $L_2$
of the BSE, Eqs.~(\protect\ref{L1L2}), are plotted versus
$q$ for $m=7~MeV$ and
$\sigma=(350~MeV)^2$. Solid lines pertain to the case $\alpha_s=0.4$, and
dashed lines to $\alpha_s=0.8$.
Starting values of the solutions are
$\varphi(0)=\pi/2$, $L_1(0)=1$, and $L_2(0)<0$.}
\label{fig:solutions}
\end{figure}

\begin{figure}
\caption{Pion mass $M_\pi$ vs. quark mass $m$ for three different
values of $\alpha_s$ ($0.0$,$0.4$,$0.8$) and $\sigma=(350~MeV)^2$.
The dotted lines essentially correspond to the normalization condition
(\protect\ref{Politzer_rs}) and $\mu=1~GeV$. The dashed lines relate to the
same scheme, but for $\mu=5~GeV$. The solid lines are obtained by using the
normalization condition (\protect\ref{Politzer_rs_mod}), slightly modified, as
explained in Sec.~\protect\ref{SDE}. The renormalization point is $\mu=1~GeV$.}
\label{fig:mp1mquark}
\end{figure}

\begin{figure}
\caption{Pion mass $M_\pi$ vs. string tension $\sigma$. As in
Fig.~\protect\ref{fig:mp1mquark}, the solid lines relate to the normalization
condition (\protect\ref{Politzer_rs_mod}).}
\label{fig:mp1sigma}
\end{figure}

\begin{figure}
\caption{Constituent mass $m^\star$ vs. quark mass $m$.
The curves are as in Fig.~\protect\ref{fig:mp1mquark}.
}
\label{fig:mc1mquark}
\end{figure}

\begin{figure}
\caption{Constituent mass $m^\star$ vs. string tension $\sigma$.
The curves are as in Fig.~\protect\ref{fig:mp1mquark}.}
\label{fig:mc1sigma}
\end{figure}

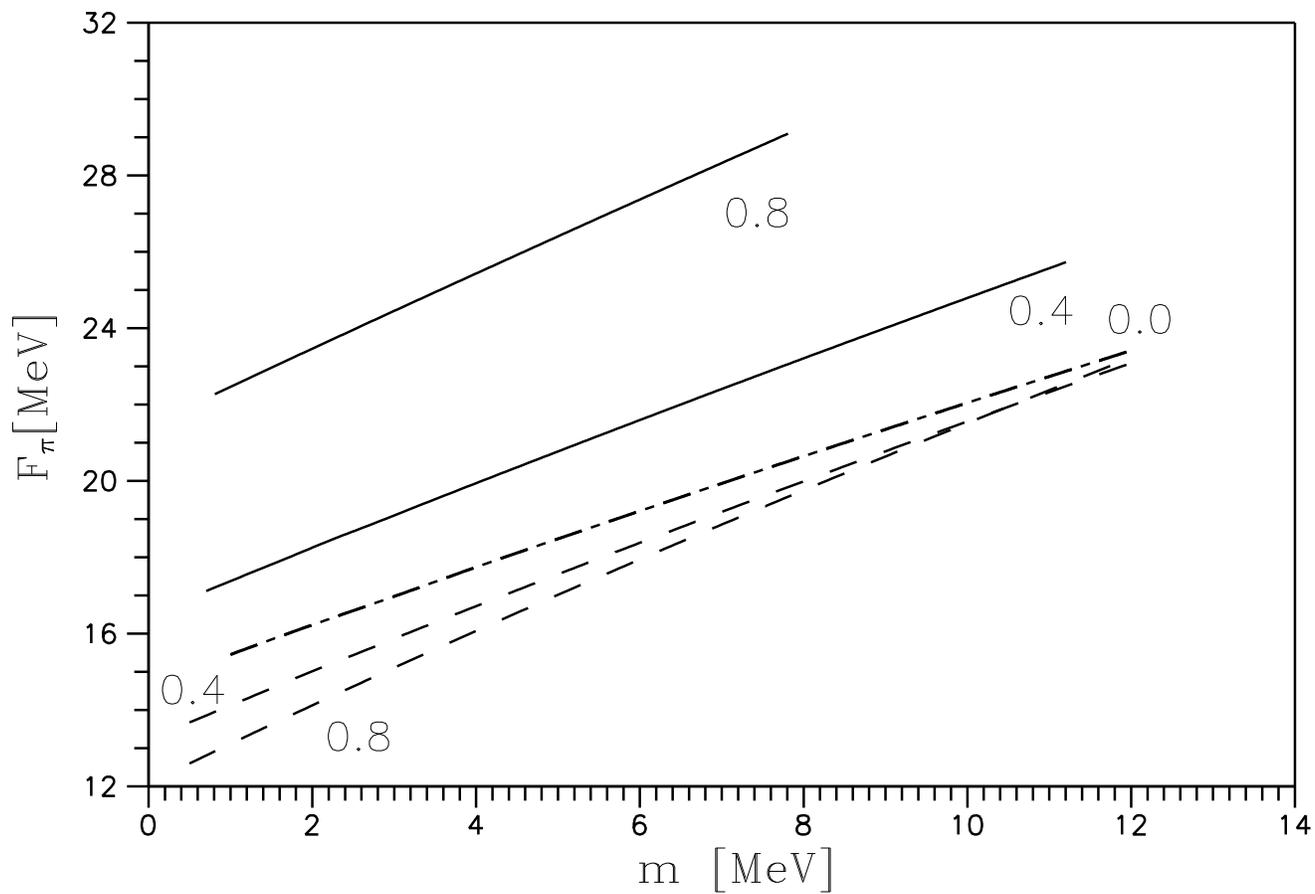
\begin{figure}
\caption{Pion decay constant $F_\pi$ vs. quark mass $m$.
The curves are as in Fig.~\protect\ref{fig:mp1mquark}.}
\label{fig:fp1mquark}
\end{figure}

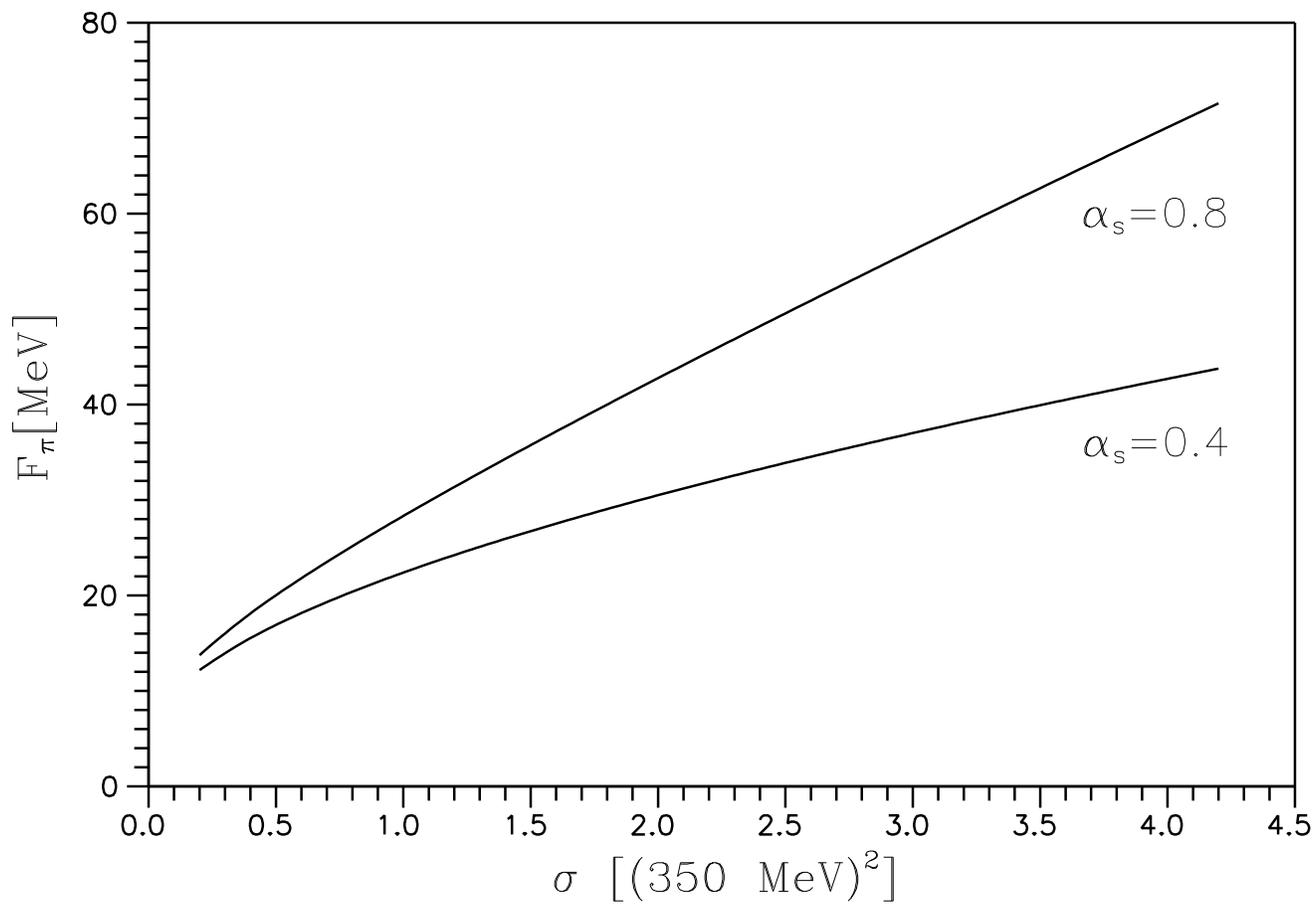
\begin{figure}
\caption{Pion decay constant $F_\pi$ vs. string tension $\sigma$.
The curves are as in Fig.~\protect\ref{fig:mp1mquark}.}
\label{fig:fp1sigma}
\end{figure}

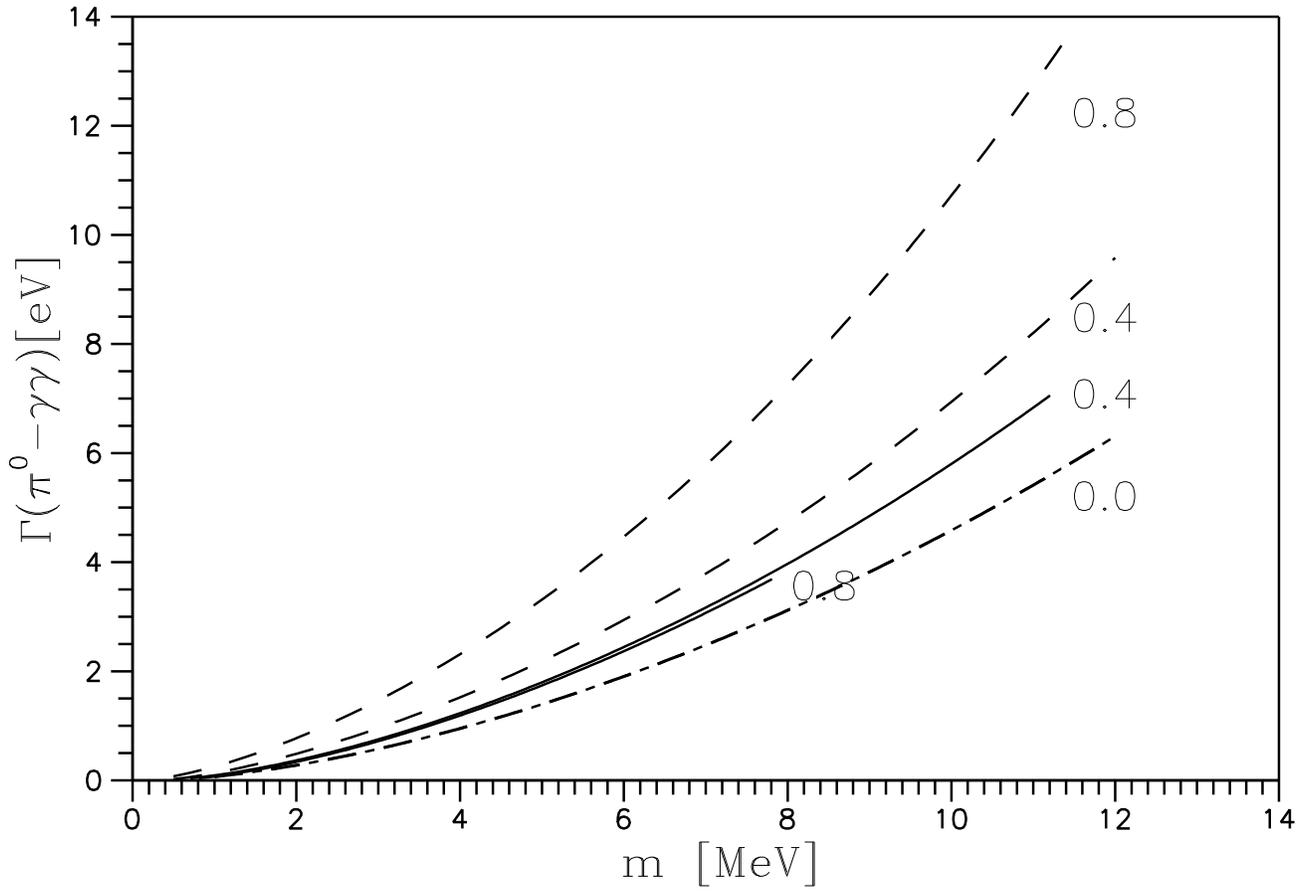
\begin{figure}
\caption{Decay width for $\pi^0\rightarrow\gamma\gamma$ vs.
quark mass $m$. The curves are as in Fig.~\protect\ref{fig:mp1mquark}.}
\label{fig:pigg1mquark}
\end{figure}

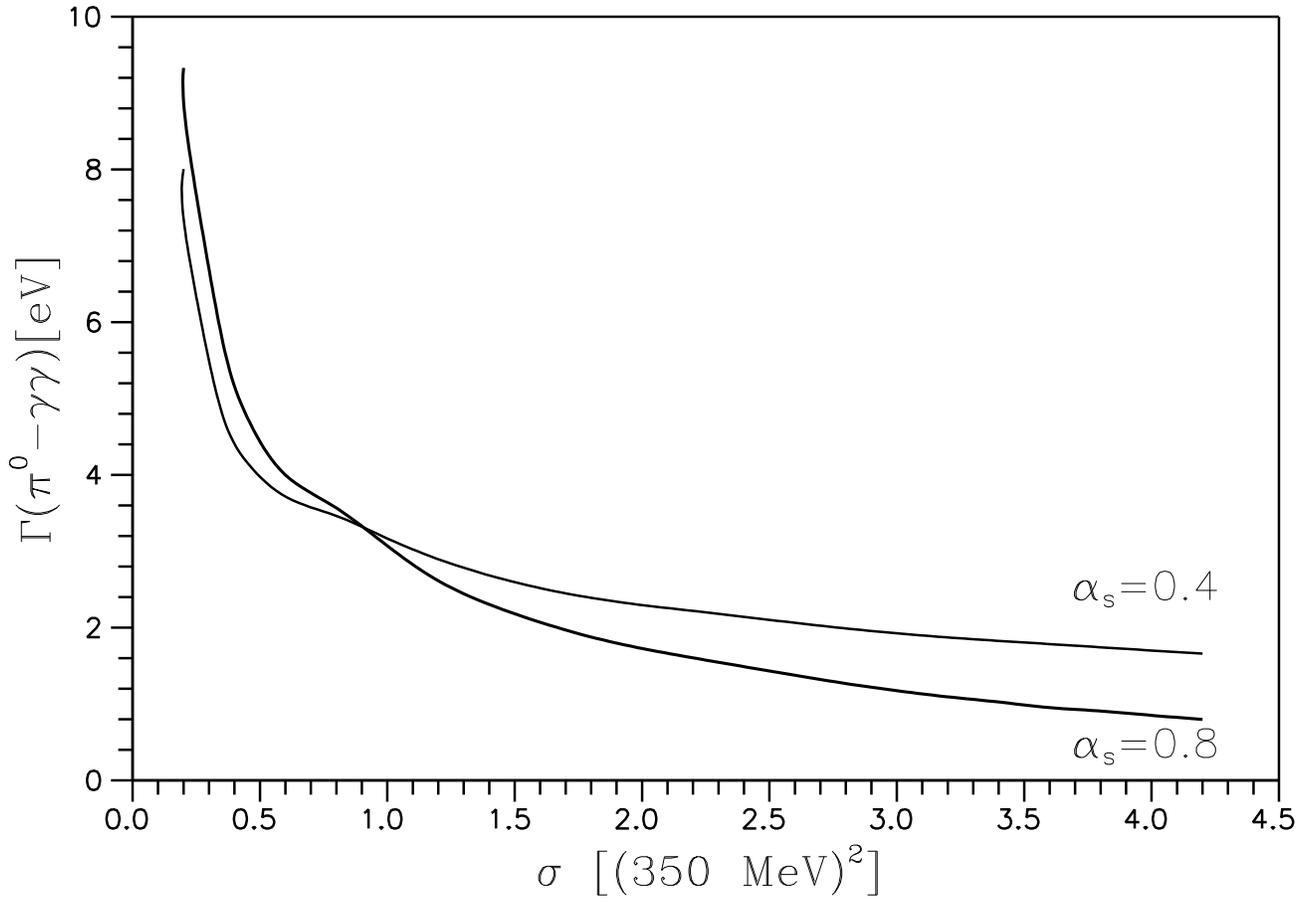
\begin{figure}
\caption{Decay width for $\pi^0\rightarrow\gamma\gamma$ vs.
string tension $\sigma$. The curves are as
in Fig.~\protect\ref{fig:mp1mquark}.}
\label{fig:pigg1sigma}
\end{figure}

\end{document}